# Self-Organization of Diverse Directional Hierarchical Networks in Simple Coupled Maps with Connection Changes

Taito Nakanishi[1], Masashi Fujii[1], Akinori Awazu[1,2,*]

[1] Graduate School of Integrated Sciences for Life, Hiroshima University, Higashi-Hiroshima, 739-8526, Japan
[2] Research Center for the Mathematics on Chromatin Live Dynamics, Hiroshima University, Higashi-Hiroshima, Hiroshima, 739-8526, Japan

**Abstract**

We comprehensively studied the morphology of the self-organized effective network structures that form in simple coupled maps with interelement synchronization-dependent connection changes. Based on the parameter values, the spontaneous formation of four types of directional hierarchical networks, named pair-driven networks, loop-driven networks, hidden trio-driven networks, and hidden community-driven networks, was observed. This study provides novel insights into the self-organized complex networks that form in neural networks, various other biological networks, and social networks.

Biological and social systems involve various self-organized network structures, such as neural networks [1-16], metabolic networks [17-20], food webs [21, 22], and human communities [23-28], which are formed through learning, adaptation, and evolution. Recently, the following coupled map system with $N$ elements that exhibit changes in the interaction strength were developed and analyzed as simple models that form complex network structures:

$$x_{n+1}^i = f((1-c)x_n^i + c \sum_{j=0}^{N-1} w_n^{ij} x_n^i) \qquad (1)$$

$$w_{n+1}^{ij} = \frac{[1+\delta g(\{x_n^i\}_{i,n})]w_n^{ij}}{\sum_{j=0}^{N-1}[1+\delta g(\{x_n^i\}_{i,n})]w_n^{ij}} \qquad (2)$$

where $x_n^i$ $(0 \leq x_n^i \leq 1)$ and $w_n^{ij}$ $(0 \leq w_n^{ij} \leq 1)$ are the state of the $i$-th element ($i = 0, 1, \ldots, N-1$) and the strength of the connection from the $j$-th element to the $i$-th one at time $n$, respectively. In addition, it is assumed that $f(x) = ax(1-x)$ (logistic map). Parameters $a$, $c$, and $\delta$ represent the logistic map, effect of the other elements, and strength of the changes in the interelement connections. Ito and Kaneko observed spontaneous differentiation between the upper- and lower-stream elements in the case of $g(\{x_n^i\}_{i,n}) = 1 - 2|x_n^i - x_n^j|$ (named the IK model) [6, 7] while Ito and Ohira observed the emergence of a pacemaker element in the case of $g(\{x_n^i\}_{i,n}) = \cos\pi(x_n^i - x_{n-1}^j)$ (named the IO model) [5]. In addition, we had previously observed a rich variety of self-organized multiple-layer networks with one-way hierarchies and/or loops, based on an exhaustive study of the IO model [16].

In this study, we focused on the $g(\{x_n^i\}_{i,n}) = \cos\pi(x_n^i - x_n^j)$ model (named the IO' model). This model was proposed in a previous study, and the phase diagrams of its parameter-dependent dynamic features, that is, $\{x_n^i\}_{i,n}$, were reported [5]. However, a detailed analysis of the network formations was not performed. In the present study, we focused on the structural features of the networks formed and observed a rich variety of hierarchical network structures in the IO' model as well; however, these structures were different from those observed in the IO model.

Figure 1 shows the phase diagrams of the typical network structures formed among the elements for a sufficiently large $n$ value in an IO' model consisting of 30 elements (N = 30) with $a = 3.6, 3.65, 3.7 \cdots, 4.0$; $c = 0.05, 0.1, \cdots, 0.25$; and $\delta = 1$. Here, we assumed that the connection from the $j$-th element to the $i$-th one exists only when $w_n^{ij} > \frac{1}{N-1}$, based on recent studies of the IK and IO models [5-7]. We found that the results were similar even for $\delta = 0.1$ and $\delta = 0.01$.

In the parameter regions marked U, G, and D, $\{w_n^{ij}\}_{i,j,n}$ exhibits the following simple features, which are similar to those observed during an analysis of the IK model [6, 7]. U: $w_n^{ij}$ were stationary ($w_{n+1}^{ij} = w_n^{ij}$) and almost uniform with $w_n^{ij} \sim \frac{1}{N-1}$; G: the elements were divided to groups such that, in each group, most of the elements were connected statically and symmetrically with each other ($w_n^{ij} = w_n^{ji}$); D: $\{w_n^{ij}\}_{i,j,n}$ exhibit disordered motion where the average values of $w_n^{ij}$ over long interval of $n$, $\langle w_n^{ij} \rangle_n$, are obtained $\sim \frac{1}{N-1}$, indicating that a specific network structure was not formed.

The behavior of the system in the case of $c \geq 0.3$ was similar to those observed in the parameter regions marked U, G, and P, as described later.

In the parameter regions marked G' and R, $\{w_n^{ij}\}_{i,j,n}$ exhibited the changes with $n$ even for large $n$ values. However, the average network structures with most of $\langle w_n^{ij} \rangle_n$ deviated sufficiently from $\frac{1}{N-1}$ and formed the following network structures. G': the elements were divided to groups such that $\langle w_n^{ij} \rangle_n = \langle w_n^{ji} \rangle_n$ in each group; R: each element was randomly connected from 10–25 elements as well as randomly connected to 10–25 elements with $\langle w_n^{ij} \rangle_n \neq \langle w_n^{ji} \rangle_n$, where the number of connections from and to each element depends on $a$ and $c$. This feature was also observed in a network named the hidden randomly connected network during a previous study on the IO model [16].

In the other parameter regions, the IO' model exhibited various network structures consisting of hierarchical directional connections among the elements. These were classified into the following four types of networks: pair-driven networks (marked P in the phase diagram), loop-driven networks (marked L), hidden trio-driven networks (marked T), and hidden community-driven networks (marked C).

The typical shape of the pair-driven networks is shown in Fig. 2(a), where $\{w_n^{ij}\}_{i,j,n}$ are stationary for large $n$ values. Some elements form pairs that are connected symmetrically with each other while the other elements form one-way directionally connected networks without any loops (a few contain branches). In addition, a few pairs are connected to the upper-most elements of the directional networks or the isolated elements. In most of the parameter regions, the pair-driven networks consist of populations of pairs and small hierarchical networks, and the number of the connections from each pair to the nonpaired elements is small. On the other hand, a large number of connections were observed from each pair for specific values of the parameters (for example, $a = 3.7$ and $c = 0.1$). In these cases, a few large networks were formed, as shown in Fig. 2(b).

The typical shape of the loop-driven networks is shown in Fig. 2(c); a few elements form a one-way connected loop while the other elements form one-way directionally connected networks without any loops (a few contain branches). In addition, some elements in the loops are connected to the upper-most elements of the directional networks. In these cases, $\{w_n^{ij}\}_{i,j,n}$ are stationary for large $n$ values, and one or two loops or networks are formed.

When hidden trio-driven or hidden community-driven networks are formed, $\{w_n^{ij}\}_{i,j,n}$ exhibit chaotic changes with $n$ even for large $n$ values. However, $\langle w_n^{ij}\rangle_n$ are not uniform but deviate sufficiently from the value $\frac{1}{N-1}$. Typical snapshots and the average shape of the trio-driven networks are shown in Fig. 3; in the average network, some of the elements form trios, which connected almost symmetrically to each other while the other elements form one-way directionally connected networks without any loops (a few contain branches). In addition, a few trios are connected to non-trio-forming elements, such that the trios are located in the upper-most stream of these directional networks. For specific values of the parameters (for example, a = 3.95 and c = 0.15), variations were observed in the trio-driven networks. For instance, networks wherein one trio was connected to another trio, and one of the elements in the trio and a few non-trio-forming elements exhibited mutual connections were observed.

Figure 4 shows typical snapshots and the average shape of the hidden community-driven networks; in the average network, some elements form a community wherein they are connected almost symmetrically to each other while the other elements form one-way directionally connected networks without any loops (a few contain branches), and each element in this community is connected to some non-community-forming elements. These networks were large hierarchical networks containing one such community in their upper-most stream.

In this study, we observed several network structures with hierarchical, directional connections among the elements in a simple dynamical system in the absence of specific rules to promote the emergence of pacemaker or leader elements, in contrast to the case for other recently proposed models [24, 26-28]. Note that the emergence of a temporally changing network showing spontaneous differentiations between the upper- and lower-stream elements was also observed in the IK model, and the underlying mechanism was reported recently by Ito and Kaneko [6, 7]. However, their detailed structural features were not investigated. We confirmed that this reported network can be classified as hidden community-driven network. In addition, we found that 1) there are two types of temporally changing dynamic networks, namely, hidden trio-driven networks and hidden community-driven networks; 2) these networks are also formed in the parameter regions that were not explored in recent studies; and 3) the hidden community-driven networks exhibit deep directional connections containing intermediate hierarchies, which form a long-term averaged network. Further, we confirmed that other network structures were also present in the IK model, although the parameter regions exhibiting loop-driven networks and hidden trio-driven networks as well as hidden community-driven networks are much narrower.

The emergence of the upper- and lower-stream elements in the hidden trio-driven and hidden community-driven networks can be explained based on mechanisms similar to those reported by Ito and Kaneko [7]. On the other hand, those in the pair-driven and loop-driven networks might be different from those for the former two network types. It should be noted that these networks were observed at the parameter values at which the attractors of the logistic map often exhibit large "windows", meaning that each element may exhibit both periodic and chaotic motions in response to small modulations in the motions of the other elements. We believe that this complexity of the motion of each element contributes to the formation of the complex networks with deep directional hierarchies. However, the elucidation of these contributions is not trivial, and a detailed analysis will be performed in a future study.

It is known that hierarchical directional connections among elements result in dramatic improvements in the learning abilities of neural network systems and are one of the characteristic features of the neuronal cell networks that form the brains of higher organisms. However, the physiological mechanisms responsible for the spontaneous formation of these hierarchical directionalities during the developmental and morphogenic processes of the brain are not well understood. It should be noted that both the IO and IO' models exhibited hierarchical directional networks; however, the networks in the IO model, which are affected by the effects of a time delay in the connection-changing rule among the elements, were formed as connected layers [16], in contrast to the case for those in the IO' model. These findings, which are based on a simple dynamical system, should help further our understanding of these processes, although further modifications of the model and a more detailed analysis are needed.

**Acknowledgments**
We thank Junji Ito and Amika Ohara for fruitful information provision.. This work was supported by Grants-in-Aid for Scientific Research (KAKENHI) (Grant Numbers

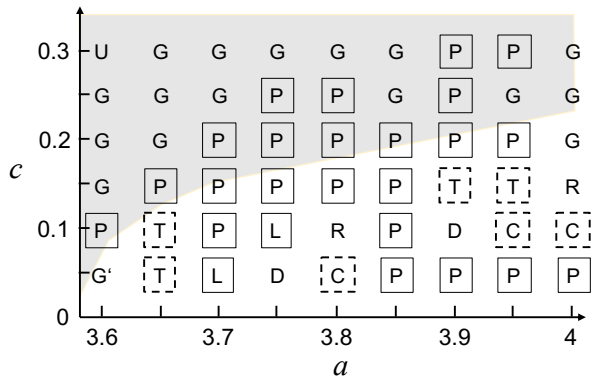

Figure 1. Phase diagram of typical network structures found in IO' model. See text for meanings of parameters and marks. Stationary and hidden hierarchical directional networks were obtained for parameters within solid and broken rectangles, respectively. Each element exhibited periodic and chaotic motions in regions with gray and white backgrounds, respectively [5].

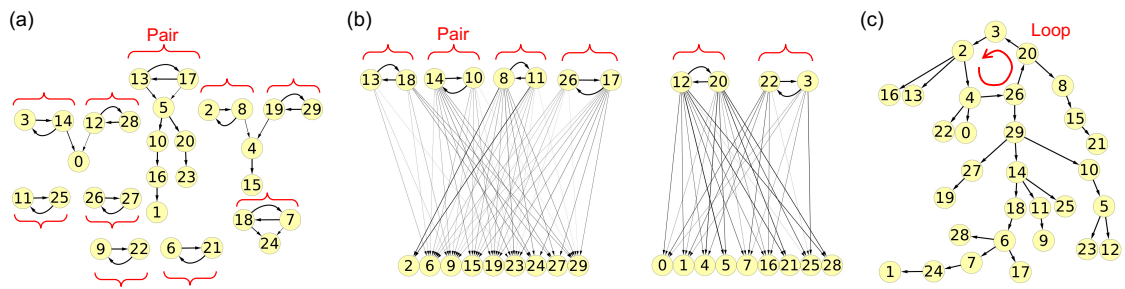

Figure 2. (a–b) Typical shapes of pair-driven networks obtained for (a) $a = 3.8$ and $c = 0.15$ and (b) $a = 3.7$ and $c = 0.1$. (c) Typical shape of loop-driven networks obtained for $a = 3.7$ and $c = 0.05$. Each circle with index $i$ indicates $i$-th element, and arrow from $i$-th circle to $j$-th one indicates existence of connection from $i$-th element to $j$-th one.

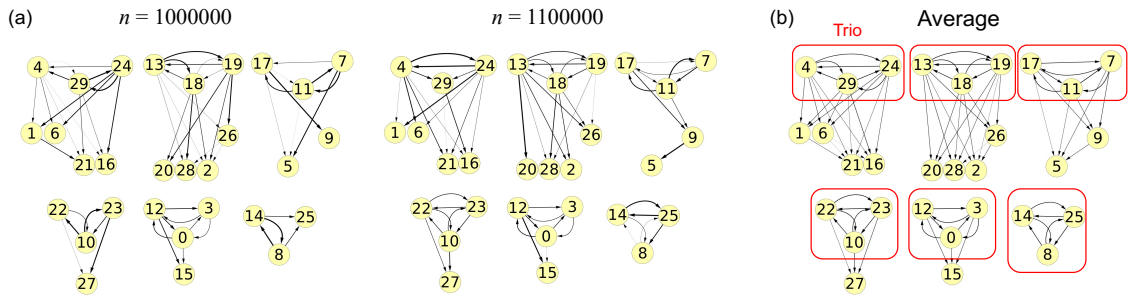

Figure 3. (a) Typical snapshots of shapes of hidden trio-driven networks at $n = 1000000$ and $n = 1100000$ for $a = 3.65$ and $c = 0.05$. (b) Averaged network structure obtained from snapshots of networks at $n = 500000–1500000$, including those in (a). Meanings of circles and arrows are the same as those in Fig. 2.

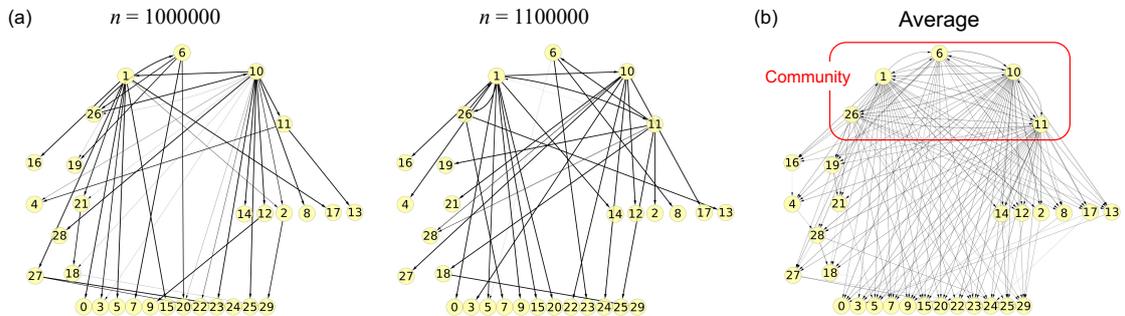

Figure 4. (a) Typical snapshots of shapes of hidden community-driven networks at $n = 1000000$ and $n = 1100000$ for $a = 4.0$ and $c = 01$. (b) Averaged network structures obtained from snapshots of networks at $n = 500000–1500000$, including those in (a). Meanings of circles and arrows are the same as those in Fig. 2